\documentclass[reprint, superscriptaddress, secnumarabic, amssymb, nobibnotes, aps, prl]{revtex4-1}

\usepackage{graphicx}
\usepackage{epstopdf}
\usepackage[T1]{fontenc}
\usepackage[latin9]{inputenc}
\usepackage{amsbsy}
\usepackage{gensymb}
\begin{document}
\title{\textrm{ Superconducting properties of the noncentrosymmetric superconductor TaOs}}
\author{D. Singh}
\affiliation{Indian Institute of Science Education and Research Bhopal, Bhopal, 462066, India}
\author{Sajilesh K. P.}
\affiliation{Indian Institute of Science Education and Research Bhopal, Bhopal, 462066, India}
\author{S. Marik}
\affiliation{Indian Institute of Science Education and Research Bhopal, Bhopal, 462066, India}
\author{A. D. Hillier}
\affiliation{ISIS facility, STFC Rutherford Appleton Laboratory, Harwell Science and Innovation Campus, Oxfordshire, OX11 0QX, UK}
\author{R. P. Singh}
\email[]{rpsingh@iiserb.ac.in}
\affiliation{Indian Institute of Science Education and Research Bhopal, Bhopal, 462066, India}

\begin{abstract}
\begin{flushleft}
\end{flushleft}
The noncentrosymmetric superconductor TaOs has been characterized using x-ray diffraction, resistivity, magnetization, and specific heat measurements. Magnetization and specific heat measurements show a bulk superconducting transition at 2.07 K. These measurements suggest that TaOs is a weakly coupled type-II superconductor. The electronic specific heat in the superconducting state can be explained by the single-gap BCS model, suggesting s-wave superconductivity in TaOs.
\end{abstract}
\maketitle
\section{Introduction}
Noncentrosymmetric (NCS) superconductivity has been studied extensively in the past few years due to their unconventional superconducting properties, which cannot be explained within the framework of BCS theory \cite{mdf,EB}. In NCS superconductor, the lack of inversion centre in the crystal structure induces an antisymmetric spin-orbit coupling (ASOC), which breaks the parity symmetry. As a result, the superconducting ground state may exhibit mixing of spin-singlet and spin-triplet components, if the pairing gap is much smaller than the strength of the spin-orbit coupling \cite{rashba,sky,kv,ia,pa, fujimoto1,fujimoto2,fujimoto3}.\\
Theoretical predictions suggested that the ratio of spin-singlet to spin-triplet pairing states in a NCS superconductor depend on the strength of ASOC. This prediction was supported by the experimental results of Li$_{2}$(Pd,Pt)$_{3}$B \cite{LPt1,LPt2,LPt3}, where pairing state was changed from spin-singlet to spin-triplet, when Pd was replaced with Pt. Concurrently, several NCS superconductors containing heavy transition elements were studied where admixed pairing states were highly anticipated due to strong spin-orbit coupling. Yet most of them showed dominant s-wave superconductivity \cite{BPS,IG,LPS1,LPS2,rw1,rw2}. In contrast, compounds with low ASOC showed unconventional superconductivity \cite{YC,LC,lnc1,lnc2}, which certainly questions the role of ASOC on the superconducting state of noncentrosymmetric superconductors.\\
Recent work on NCS superconductors are predominantly focused on compounds with $\alpha-Mn$ structure after the discovery of time-reversal symmetry (TRS) breaking in Re$_{6}$Zr \cite{rz1}. In this system, Re-atom occupies all the noncentrosymmetric sites, therefore was considered a worthy candidate to study the effects of lack of inversion symmetry on the superconducting state. Instigated by the above finding, several other Re-based compounds were systematically investigated, where the transition metal element with Re was replaced with other heavier elements to tune the strength of ASOC e.g. Re$_{24}$Ti$_{5}$ \cite{RT1}, Nb$_{0.18}$Re$_{0.82}$ \cite{nr1,nr2}, and Re$_{6}$Hf \cite{rf,rhf}. Most of these compounds exhibited single-band superconductivity except Nb$_{0.18}$Re$_{0.82}$, which showed double-gap superconductivity \cite{nr2}. Hence, no apparent conclusion can be made of the role of ASOC in determining the pairing state of noncentrosymmetric superconductors.\\
Another compound with $\alpha-Mn$ structure which we studied recently is Nb$_{0.5}$Os$_{0.5}$ \cite{NB}, which shows s-wave superconductivity when examined from bulk and $\mu$SR measurements. In order to address the question regarding the effects of ASOC, we have replaced Nb with Ta, as Ta atom is heavier than Nb atom, substituting it should enhance the strength of spin-orbit coupling which, in turn, can increase the extent of parity mixing in the superconducting ground state.\\ 
	In this work, we report the detailed characterization of the noncentrosymmetric superconductor TaOs exhibiting bulk superconductivity at $T_{c}$ = 2.07 K. Superconducting properties were determined by the magnetic susceptibility, electrical resistivity, and specific heat measurements. The results indicate a single-gap s-wave superconductivity with negligible effect of enhanced ASOC.
\section{Experimental Details}
The sample of TaOs was prepared by melting stoichiometric amounts of Ta (99.95$\%$, Alfa Aesar) and Os (99.95$\%$, Alfa Aesar) in an arc furnace. The ingot was flipped and remelted several times. The observed weight loss during the melting was negligible. Then, the ingot was annealed in a vacuum-sealed quartz tube at 900 $^{\circ}$C for 1 week. followed by cooling to room temperature in 24 hours. The powder x-ray diffraction (XRD) spectrum was collected on a X'pert PANalytical diffractometer. The magnetization measurements were performed using superconducting quantum interference device (SQUID, Quantum Design Inc.) and electrical resistivity and specific heat measurements were done in a physical property measurement system (PPMS, Quantum Design Inc.).
\begin{figure}[h]
\includegraphics[width=1.0\columnwidth]{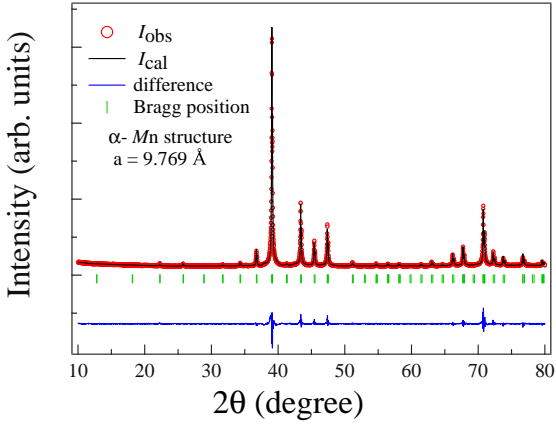}
\caption{\label{Fig1:xrd} Powder XRD pattern for the TaOs sample recorded at room temperature.}
\end{figure}
\section{Results and Discussion}

The x-ray diffraction pattern of TaOs shown in Fig. 1. No impurities were observed in the diffraction pattern. Rietveld refinement performed on the sample confirms that the sample crystallizes into cubic, noncentrosymmetric $\alpha$ - $Mn$ structure (space group $I \bar{4}3m$, No. 217) with the lattice cell parameter a = b = c = 9.769 $\pm$ 0.002 \text{\AA}, which is in good agreement with the published data \cite {PS}.\\ 
Figure 2(a) shows the electrical resistivity data as a function of temperature in the range of 1.8 K $\le$ $\textit{T}$ $\le$ 300 K in zero applied magnetic field. The measurement shows that the sample has poor metallic behavior. This is similar to other $\alpha-Mn$ structure compounds \cite{RT1,rw1,nr1,rhf, NB} where the similar behaviour was attributed to electron scattering due to disorder. The resistivity drops to zero at $T_{c}$ $\simeq$ 2.06 K as shown in the inset of Fig.2(a).\\
\begin{figure}
\includegraphics[width=1.0\columnwidth]{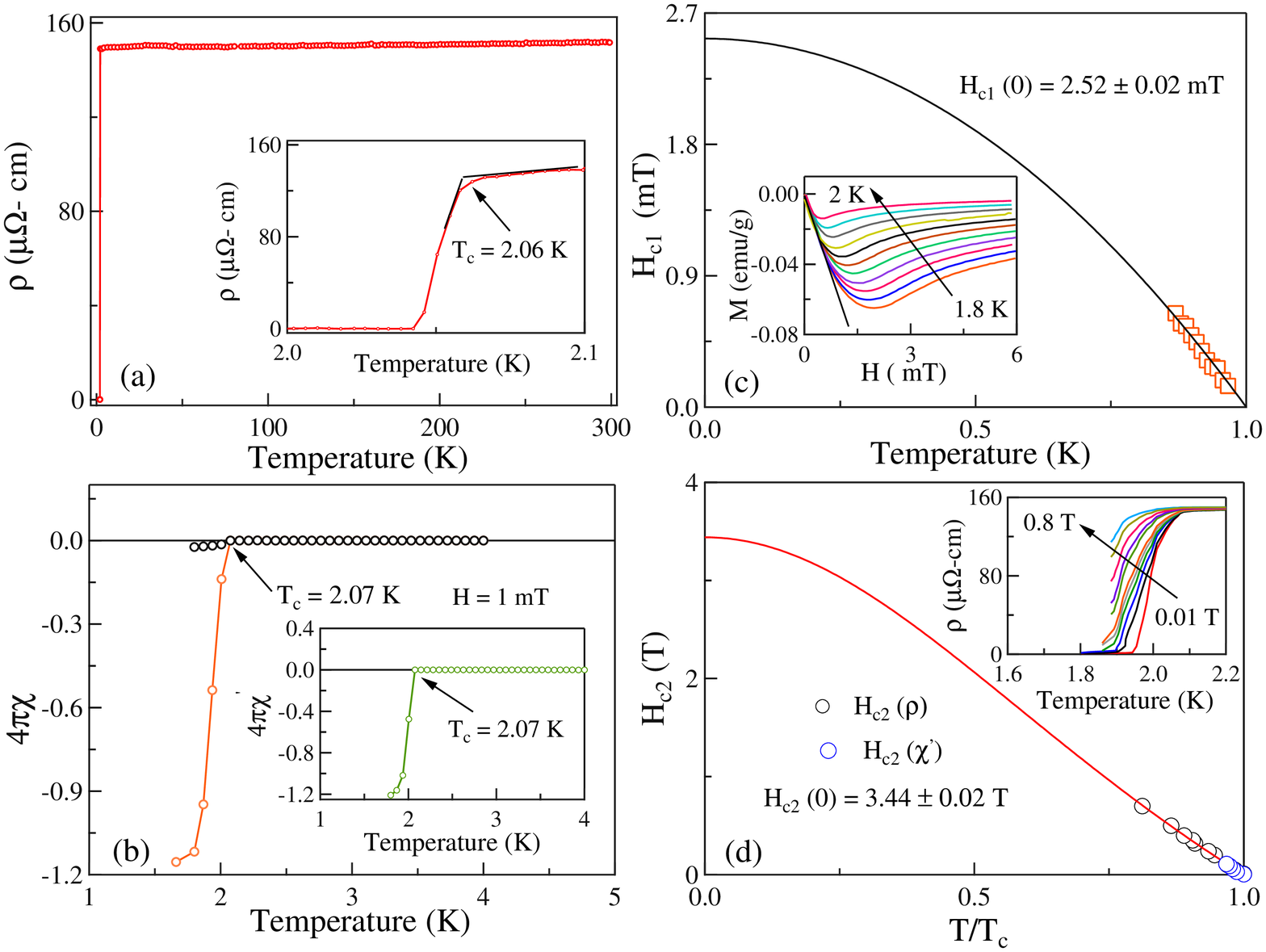}
\caption{\label{Fig2:Resis} a) Temperature dependence of electrical resistivity
for TaOs between 1.8 and 300 K taken in zero magnetic
field. Inset highlights the superconducting transition temperature
at  $T_{c}$ = 2.06 K. (b) The magnetization measurement for TaOs taken in 1 mT field showing superconducting transition at $T_{c}^{onset}$ = 2.07 K. (c) Temperature dependence of lower critical field yields $H_{c1}$(0) = 2.52 $\pm$ 0.02 mT. The inset shows the M(H) curves taken at different fixed temperatures. (d) Temperature dependence of upper critical field $H_{c2}$(T) yields $H_{c2}$(0) = 3.44 $\pm$ 0.02 T. Inset shows the resistivity measurements at different fixed applied magnetic fields.}
\end{figure}
The magnetization measurement was performed in an applied field of H = 1 mT, confirms bulk superconductivity with the onset of strong diamagnetic signal around $T_{c}^{onset}$ = 2.07 K, as displayed in Fig. 2(b). The superconducting volume fraction is little higher than -1, that may be due to demagnetization effect. The $M(H)$ curve obtained for temperature above $T_{c}$ (not shown here) was almost linear in $H$, which when fitted with the linear model yields the intrinsic susceptibility $\chi$ = 8.81 $\times$ 10$^{-4}$ cm$^{3}$/mol. The measured susceptibility $\chi$ results from the susceptibilities from the core and conduction electrons and given by $\chi$  = $\chi_{core}$ + $\chi_{vv}$ + $\chi_{L}$ + $\chi_{P}$, where $\chi_{core}$ is the diamagnetic core susceptibility, $\chi_{vv}$ the paramagnetic Van Vleck susceptibility, $\chi_{L}$ the Landau diamagnetic susceptibility, and $\chi_{P}$ the Pauli spin susceptibility. Here the $\chi_{core}$ contribution is from the core electrons, whereas $\chi_{L}$ , $\chi_{P}$ is due to conduction electrons. Using the diamagnetic susceptibilities of the constituent elements \cite{core} gives $\chi_{core}$ = $\text{-}$ 6.05$\times$ 10$^{-5}$ cm$^{3}$/mol. The $\chi_{P}$ is given by $\chi_{P}$ = (g$^{2}$/4)$\mu$$^{2}_{B}$$D(E_{F}$) \cite{ad}, where g is the spectroscopic splitting factor of the conduction carriers, $\mu_{B}$ the Bohr magneton, and $D(E_{F}$) the band-structure density of states at the Fermi energy E$_{F}$. Using g = 2 and $D(E_{F}$) = 1.27 states/eV f.u. (estimated from specific heat measurements), we get $\chi_{P}$ = 4.11 $\times$ 10$^{-5}$ cm$^{3}$/mol. Taking the band structure effective mass $m^{*}_{band}$ = m$_{e}$, where $m_{e}$ is the mass of a free electron, we obtained $\chi_{L}$ = $\text{-}$1.37 $\times$ 10$^{-5}$ cm$^{3}$/mol from the formula $\chi_{L}$ = $\text{-}$1/3($m_{e}$/$m^{*}_{band}$)$^{2}$$\chi_{P}$ \cite{sr}. Using the above- estimated values $\chi_{vv}$ is derived as 9.14 $\times$ 10$^{-5}$ cm$^{3}$/mol.

	To calculate the lower critical field $H_{c1}$(0), magnetization curves $M(H)$ in low applied magnetic fields was measured at various temperatures from 1.8 K to 2 K as shown in the inset of Fig. 2(c). The lower critical field $H_{c1}$ is defined as the point at which the magnetization deviates from linearity. The main panel of Fig. 2(c) shows the temperature variation of $H_{c1}$(T), which can be described by the formula
\begin{equation}  
H_{c1}(T) = H_{c1}(0)\left(1-\left(\frac{T}{T_{c}}\right)^{2}\right).
\label{eqn1:hc1}
\end{equation}

When fitted to the experimental data, it yields $H_{c1}$(0) = 2.52 $\pm$ 0.02 mT.\\
The temperature dependence of the upper critical field $H_{c2}$(T) was determined by measuring the shift in $T_{c}^{mid}$ in different fixed applied magnetic fields in resistivity measurements as shown in Fig. 2(d). It is evident from the graph that the data obtained from the measurements vary linearly with temperature. This data can be fitted using the relation given by
\begin{equation}
H_{c2}(T) = H_{c2}(0)\frac{(1-t^{2})}{(1+t^2)},
\label{eqn2:hc2}
\end{equation} 
where t = $T/T_{c}$. By fitting above equation in the $H_{c2}$-T graph, it yields $H_{c2}$(0) $\simeq$ 3.44 $\pm$ 0.02 T.
$H_{c2}$(0) can be used to estimate the Ginzburg Landau coherence length $\xi_{GL}$ from the relation \cite{mtin} 
\begin{equation}
H_{c2}(0) = \frac{\Phi_{0}}{2\pi\xi_{GL}^{2}} ,
\label{eqn3:up}
\end{equation}                                                                                      
where $\Phi_{0}$ is the quantum flux $(h/2e)$. For $H_{c2}$(0) $\simeq$ 3.44 $\pm$ 0.02 T, we obtained $\xi_{GL}(0)$ = 97.9 $\pm$ 0.3 \text{\AA}.
Within the $\alpha$-model the Pauli limiting field is given by \cite{dc}
\begin{equation}
H_{c2}^{p}(0) = 1.86T_{c}\left(\frac{\alpha}{\alpha_{BCS}}\right) .
\label{eqn4:pauli}
\end{equation}
Using $\alpha$ = 1.71 (from the specific heat measurement), it yields $H_{c2}^{p}$(0) = 3.73 T. The upper critical field $H_{c2}$(0) and Pauli limiting field are close. Detailed investigation in low temperatures$/$single crystals are required to calculate accurate value of $H_{c2}$(0) and confirm the contribution of spin-triplet component in superconducting ground state. 
The Ginzburg Landau penetration depth $\lambda_{GL}$(0) can be obtained from the $H_{c1}$(0) and $\xi_{GL}$(0) using the relation \cite{mtin}
\begin{equation}
H_{c1}(0) = \frac{\Phi_{0}}{4\pi\lambda_{GL}^2(0)}\left(\mathrm{ln}\frac{\lambda_{GL}(0)}{\xi_{GL}(0)}+0.12\right) .  
\label{eqn5:ld}
\end{equation} 
Using $H_{c1}$(0) = 2.52 $\pm$ 0.02 mT and $\xi_{GL}$(0) = 97.9 $\pm$ 0.3 $\text{\AA}$, we obtained $\lambda_{GL}$(0) $\simeq$ 5168 $\pm$ 3 $\text{\AA}$. 
The Ginzburg Landau parameter is given by the relation \cite{mtin}
\begin{equation}
\kappa_{GL} = \frac{\lambda_{GL}(0)}{\xi_{GL}(0)} .
\label{eqn6:kgl}
\end{equation}
For $\xi_{GL}$(0) = 97.9 $\pm$ 0.3 $\text{\AA}$ and $\lambda_{GL}$(0) = 5168 $\pm$ 3 $\text{\AA}$, we calculated $\kappa_{GL}$ $\simeq$ 52.78 $\pm$ 0.13. This indicates type-II superconductivity in TaOs.
Thermodynamic critical field $H_{c}$ can be estimated from $\kappa_{GL}$(0) and $H_{c2}$(0) using the relation 
\begin{equation}
H_{c} = \frac{H_{c2}}{\sqrt{2}\kappa_{GL}} ,
\label{eqn7:k}
\end{equation}                                                                                 
which for $H_{c2}$ = 3.44 $\pm$ 0.02 T and $\kappa_{GL}$ = 52.78 $\pm$ 0.13 yields $H_{c}$ = 46.09 $\pm$ 0.15 mT.\\ 
\begin{figure}
\includegraphics[width=1.0\columnwidth]{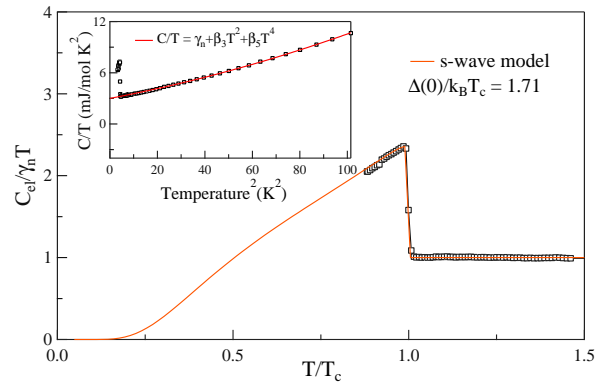}
\caption{\label{Fig3:cp} The low temperature specific heat data in the superconducting regime fitted with the single gap s-wave model using Eq. 11 for $\Delta$(0)/$k_{B}$T = 1.71. Inset: The C/T vs T$^{2}$ data in the temperature range 3 K $\le$ $\textit{T}$ $\le$ 100 K, fitted with low temperature Debye model.} 
\end{figure}

	The low temperature specific heat measurement $C(T)$ was taken in zero applied field. The specific heat data in Fig. 3 confirms bulk superconductivity in TaOs. The normal state low temperature specific heat data above $T_{c}$ was fitted with the relation 
	\begin{equation}
\frac{C}{T} = \gamma_{n}+\beta_{3}T^{2}+\beta_{5}T^{4} ,
\label{eqn8:cpp}
\end{equation}
where $\gamma_{n}$ is the normal state Sommerfeld coefficient related to the electronic contribution to the specific heat whereas $\beta_{3}$ and $\beta_{5}$ are the coefficients related to the lattice contribution to the specific heat. The solid red line in inset of Fig. 3 shows the best fit to the data which yields $\gamma_{n}$ = 3.0 $\pm$ 0.01 mJ mol$^{-1}$ K$^{-2}$, $\beta_{3}$ = 0.052 $\pm$ 0.002 mJ mol$^{-1}$ K$^{-4}$, and $\beta_{5}$ = 0.22 $\pm$ 0.04 $\mu$J mol$^{-1}$ K$^{-6}$. The Debye temperature was related to the coefficient $\beta_{3}$ which give $\theta_{D}$ = 332 K. Density of states at the Fermi level $D_{C}(E_{F})$ was estimated 1.27 $\frac{states}{eV f.u}$ using the relation $\gamma_{n}$ = $(\pi^{2}k_{B}^{2}/3)D_{C}(E_{F})$.\\
The electron-phonon coupling constant which gives the strength of the attractive interaction between the electron and phonon can be calculated by the McMillan equation \cite{WL},
\begin{equation}
\lambda_{e-ph} = \frac{1.04+\mu^{*}\mathrm{ln}(\theta_{D}/1.45T_{c})}{(1-0.62\mu^{*})\mathrm{ln}(\theta_{D}/1.45T_{c})-1.04 } ,
\label{eqn9:ld}
\end{equation}                       
where $\mu^{*}$ = 0.13 is the Coulomb repulsion parameter. Using $T_{c}$ = 2.07 K and $\theta_{D}$ = 332 K for TaOs, we obtained $\lambda_{e-ph}$ $\simeq$ 0.50. This value suggests that TaOs is a weakly coupled superconductor.\\ 
Using the electron-phonon coupling constant we can calculate the bare-band effective mass $m^{*}$ of the quasi-particles which contains the influence of many body electron-phonon interactions, which for $\lambda_{e-ph}$ = 0.50, gives (assuming $m^{*}_{band}= m_{e}$) $m^{*}$ = 1.50 $m_{e}$ \cite{GG}.\\
The electronic contribution to the specific heat can be calculated by subtracting the phononic contribution from the total specific heat. The normalized specific heat jump $\frac{\Delta C_{el}}{\gamma_{n}T_{c}}$ is 1.41 for $\gamma_{n}$ = 3.0 mJ mol$^{-1}$ K$^{-2}$. The value obtained for $\frac{\Delta C_{el}}{\gamma_{n}T_{c}}$ is close to the value for a weakly coupled BCS type superconductor ( = 1.43).\\
The temperature dependence of the normalized entropy S in the superconducting state for a single-gap BCS superconductor is given by 
\begin{equation}
\frac{S}{\gamma_{n}T_{c}} = -\frac{6}{\pi^2}\left(\frac{\Delta(0)}{k_{B}T_{c}}\right)\int_{0}^{\infty}[ \textit{f}\ln(f)+(1-f)\ln(1-f)]dy ,
\label{eqn10:s}
\end{equation}
where $\textit{f}$($\xi$) = [exp($\textit{E}$($\xi$)/$k_{B}T$)+1]$^{-1}$ is the Fermi function, $\textit{E}$($\xi$) = $\sqrt{\xi^{2}+\Delta^{2}(t)}$, where $\xi$ is the energy of normal electrons measured relative to the Fermi energy, $\textit{y}$ = $\xi/\Delta(0)$, $\mathit{t = T/T_{c}}$, and $\Delta(t)$ = tanh[1.82(1.018(($\mathit{1/t}$)-1))$^{0.51}$] \cite{BM} is the BCS approximation for the temperature dependence of the energy gap. The normalized electronic specific heat is related to the normalized entropy by
\begin{equation}
\frac{C_{el}}{\gamma_{n}T_{c}} = t\frac{d(S/\gamma_{n}T_{c})}{dt} ,
\label{eqn11:Cel}
\end{equation}
where $C_{el}$ below $T_{c}$ is described by Eq. (11) whereas above $T_{c}$ its equal to $\gamma_{n}T_{c}$. The specific heat data in Fig. 3 fits perfectly well for a fitting parameter $\alpha$ = $\Delta(0)/k_{B}T_{c}$ = 1.71 $\pm$ 0.02, which is close to the BCS value $\alpha_{BCS}$ = 1.764 in the weak coupling limit, suggesting that TaOs have dominant s-wave superconductivity.\\
Recently unconventional vortex dynamics have been observed in some noncentrosymmetric superconductors \cite{MS,ED,CF1,CF2}, which is very distinct from the classical and high-$T_{c}$ superconductors. Therefore, it is necessary to measure the stability of the vortex system against the thermal fluctuations which is given by Ginzburg number $G_{i}$. Ginzburg number $G_{i}$ is the ratio of thermal energy $k_{B}$$T_{c}$ to the condensation energy associated with coherence volume \cite{vortex}
\begin{equation}
 G_{i} = \frac{1}{2}\left(\frac{k_{B}\mu_{0}\tau T_{c}}{4\pi\xi^{3}(0)H_{c}^{2}(0)}\right)^2 .
\label{eqn12:gi}
\end{equation}                                                                          
Here $\tau$ is the anisotropy parameter which is 1 for cubic TaOs. For $\xi$(0) = 97.9  \text{\AA}, $H_{c}$(0) = 46.09 mT and $T_{c}$= 2.07 K, we got $G_{i}$= 1.02 $\times$ 10$^{-6}$. The value of $G_{i}$ is more towards the low $T_{c}$ superconductors ($G_{i}$ $\simeq$ 10$^{-8}$), suggesting that thermal fluctuations may not be playing any important role in vortex unpinning in our system.\\
Uemura $\textit{et al.}$ identified that the class of a superconductor can be differentiated conveniently based on the ratio of the transition temperature ($T_{c}$) to the Fermi temperature ($T_{F}$) \cite{YJU}. It was shown that unconventional, exotic superconductors fall in the range of 0.01 $\leq$ $\frac{T_{c}}{T_{F}}$ $\leq$ 0.1.

\begin{table}[h!]
\caption{Superconducting and normal properties of TaOs}
\begin{center}
\begin{tabular}[b]{lccc}\hline\hline
Parameters& unit& TaOs\\
\hline
\\[0.5ex]                                  
$T_{c}$& K& 2.07\\             
$H_{c1}(0)$& mT& 2.52\\
$H_{c}(0)$& mT& 46.09\\                       
$H_{c2}(0)$& T& 3.44\\
$H_{c2}^{P}(0)$& T& 3.73\\
$\xi_{GL}$& \text{\AA}& 97.9\\
$\lambda_{GL}$& \text{\AA}& 5168\\
$k_{GL}$& &52.78\\
$\Delta C_{el}/\gamma_{n}T_{c}$&   &1.41\\
$\Delta(0)/k_{B}T_{c}$&  &1.71\\
$m^{*}/m_{e}$& & 12.1\\             
n& 10$^{27}$m$^{-3}$& 4.1\\
$l$&  \text{\AA}& 33.2\\ 
$\xi_{0}$&  \text{\AA}& 74.03\\                      
$\xi_{0}/l$& & 2.23\\
$v_{f}$& 10$^{4}$ms$^{-1}$& 4.74\\
$\lambda_{L}$& \text{\AA}& 2875.5\\
$T_{c}$/$T_{F}$& &0.0023\\
\\[0.5ex]
\hline\hline
\end{tabular}
\par\medskip\footnotesize
\end{center}
\end{table}

For a 3D system Fermi temperature T$_{F}$ is given by the relation
\begin{equation}
 k_{B}T_{F} = \frac{\hbar^{2}}{2}(3\pi^{2})^{2/3}\frac{n^{2/3}}{m^{*}}, 
\label{eqn13:tf}
\end{equation}
where n is the quasiparticle number density per unit volume. 
Using the Sommerfeld coefficient for TaOs, we can calculate the quasiparticle number density per unit volume and mean free path \cite{ck}
\begin{equation}
\gamma_{n} = \left(\frac{\pi}{3}\right)^{2/3}\frac{k_{B}^{2}m^{*}V_{\mathrm{f.u.}}n^{1/3}}{\hbar^{2}N_{A}}
\label{eqn14:gf}
\end{equation}
where k$_{B}$ is the Boltzmann constant, N$_{A}$ is the Avogadro constant, V$_{\mathrm{f.u.}}$ is the volume of a formula unit and m$^{*}$ is the effective mass of quasiparticles. The electronic mean free path $\textit{l}$ is related to residual resistivity $\rho_{0}$ by the equation
 \begin{equation}
\textit{l} = \frac{3\pi^{2}{\hbar}^{3}}{e^{2}\rho_{0}m^{*2}v_{\mathrm{F}}^{2}}
\label{eqn15:le}
\end{equation}
where the Fermi velocity $v_{\mathrm{F}}$ is related to the effective mass and the carrier density by
\begin{equation}
n = \frac{1}{3\pi^{2}}\left(\frac{m^{*}v_{\mathrm{f}}}{\hbar}\right)^{3} .
\label{eqn16:n}
\end{equation}
In the dirty limit, the penetration depth $\lambda_{GL}$(0) can be estimated by relation
\begin{equation}
\lambda_{GL}(0) = \lambda_{L}\left(1+\frac{\xi_{0}}{\textit{l}}\right)^{1/2}
\label{eqn17:f}
\end{equation}
where $\xi_{0}$ is the BCS coherence length. The $\lambda_{L}$ is the London penetration depth, which is given by
\begin{equation}
\lambda_{L} = \left(\frac{m^{*}}{\mu_{0}n e^{2}}\right)^{1/2}
\label{eqn18:laml}
\end{equation}
The Ginzburg-Landau coherence length is also affected in the dirty limit. The relationship between the BCS coherence length $\xi_{0}$ and the Ginzburg-Landau coherence $\xi_{GL}$(0) at T = 0 is
\begin{equation}
\frac{\xi_{GL}(0)}{\xi_{0}} = \frac{\pi}{2\sqrt{3}}\left(1+\frac{\xi_{0}}{\textit{l}}\right)^{-1/2}
\label{eqn19:xil}
\end{equation}
Equations (14)-(19) form a system of four equations which can be used to estimate the parameters m$^{*}$, n, $\textit{l}$, and $\xi_{0}$ as done in Ref.\cite{DAM}. The system of equations was solved simultaneously using the values $\gamma_{n}$ = 3.0 mJ mol$^{-1}$K$^{-2}$, $\xi_{GL}$(0) = 97.9 \text{\AA}, and $\rho_{0}$ = 150.01 $\mu$$\Omega$-cm. The estimated values are tabulated in Table 1. It is clear that $\xi_{0}$ > $\textit{l}$, indicating that TaOs is in the dirty limit. The estimated value of mean free path $\textit{l}$ is of the same order as observed in other $\alpha$ - $Mn$ structure noncentrosymmetric superconductors, where similar high residual resistivity and dirty limit superconductivity was observed \cite{rw1,nr2,DAM}. 
\begin{figure}
\includegraphics[width=1.0\columnwidth]{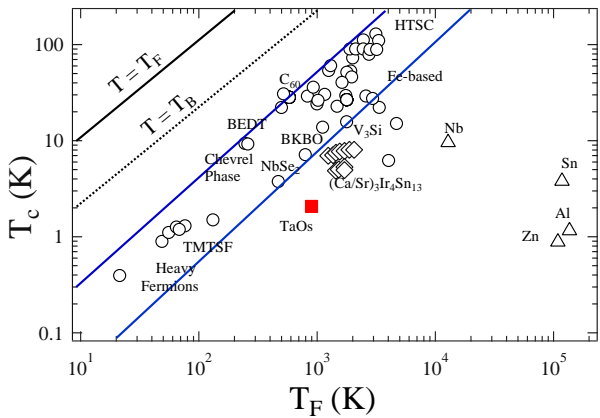}
\caption{\label{Fig4:up} The Uemura plot showing the superconducting transition temperature $T_{c}$ vs the effective Fermi temperature $T_{F}$, where TaOs is shown as a solid red square. Other data points plotted between the blue solid lines is the different families of unconventional superconductors \cite{KKC,RKH}.} 
\end{figure}
	
	Using the estimated value of n in Eq. (13) we get $T_{F}$ = 897 K, giving $\frac{T_{c}}{T_{F}}$ = 0.0023, which places TaOs away from the unconventional superconductors as shown by a solid red square in Fig. 4, where blue solid lines represent the band of unconventional superconductors.\\
\section{Conclusion}
In summary, TaOs was prepared by standard arc melting technique. The noncentrosymmetric $\alpha$-$Mn$ cubic structure was confirmed by XRD analysis. A comprehensive study of the superconducting properties of TaOs was done using resistivity, magnetic susceptibility, and heat capacity measurements. These measurements suggest type-II superconductivity in TaOs with superconducting transition temperature $T_{c}$ = 2.07 K. The electronic specific heat in the superconducting state is well described by the single-gap BCS expression, suggesting the $\textit{s}$-wave superconductivity. The close value of the upper critical field $H_{c2}$(0) and Pauli limiting field in noncentrosymmetric superconductors may indicate the possibility of mixed pairing in the superconducting ground state. In order to confirm it, local probe measurements, e.g. muon spin rotation/ relaxation is vital.\\

\section{Acknowledgments}

R.~P.~S.\ acknowledges Science and Engineering Research Board, Government of India for the Young Scientist Grant YSS/2015/001799 and DST FIST.

\end{document}